\documentstyle[12pt,epsf]{article}

\begin{document}

\baselineskip=18.8pt plus 0.2pt minus 0.1pt

\makeatletter

\@addtoreset{equation}{section}
\renewcommand{\theequation}{\thesection.\arabic{equation}}
\renewcommand{\thefootnote}{\fnsymbol{footnote}}
\newcommand{\nn}{\nonumber}
\newcommand{\bm}[1]{\mbox{\boldmath $#1$}}
\newcommand{\Half}{\frac{1}{2}}
\newcommand{\half}{1/2}
\newcommand{\B}{{\cal B}} 
\newcommand{\E}{{\cal E}} 
\newcommand{\X}{X}        
\newcommand{\x}{x}        
\newcommand{\Y}{Y}        
\newcommand{\y}{y}        
\newcommand{\A}{A}        
\renewcommand{\u}{u}      
\renewcommand{\v}{v}      
\newcommand{\g}{g}        
\newcommand{\e}{e}        
\newcommand{\p}{\partial}
\newcommand{\tr}{\mathop{\rm Tr}}
\newcommand{\diag}{\mathop{\rm diag}}
\newcommand{\r}{r}
\newcommand{\hr}{\hat{\r}}
\newcommand{\bmr}{\bm{\r}}
\newcommand{\hbmr}{\hat{\bm{\r}}}
\newcommand{\Pdrv}[2]{\frac{\p #1}{\p #2}}
\newcommand{\pdrv}[2]{{\p #1}/{\p #2}}
\newcommand{\calD}{{\cal D}}
\newcommand{\com}[2]{\left[#1,#2\right]}
\newcommand{\barm}{\overline{m}}
\newcommand{\vp}{\varphi}
\newcommand{\nr}{\varphi_+}
\newcommand{\pr}{\varphi_-}
\newcommand{\D}{D}
\newcommand{\Drv}[2]{\frac{d #1}{d #2}}
\newcommand{\wt}[1]{\widetilde{#1}}

\makeatother

\begin{titlepage}
\title{
\hfill\parbox{4cm}
{\normalsize KUNS-1499\\HE(TH)~98/03\\{\tt hep-th/9803127}}\\
\vspace{1cm}
3-String Junction and BPS Saturated Solutions\\
in $SU(3)$ Supersymmetric Yang-Mills Theory
}
\author{
Koji {\sc Hashimoto}\thanks{{\tt hasshan@gauge.scphys.kyoto-u.ac.jp}.
Supported in part by Grant-in-Aid for Scientific
Research from Ministry of Education, Science and Culture
(\#3160).},
{}
Hiroyuki {\sc Hata}\thanks{{\tt hata@gauge.scphys.kyoto-u.ac.jp}.
Supported in part by Grant-in-Aid for Scientific
Research from Ministry of Education, Science and Culture
(\#09640346).}
{} and
Naoki {\sc Sasakura}\thanks{{\tt sasakura@gauge.scphys.kyoto-u.ac.jp}}
\\[7pt]
{\it Department of Physics, Kyoto University, Kyoto 606-8502, Japan}
}
\date{\normalsize March, 1998}
\maketitle
\thispagestyle{empty}

\begin{abstract}
\normalsize
We construct BPS saturated regular configurations of ${\cal N}=4$
$SU(3)$ supersymmetric Yang-Mills theory carrying
non-parallel electric and magnetic charges.
These field theory BPS states correspond to  the string theory
BPS states of 3-string junctions
connecting three different D3-branes by regarding
the ${\cal N}=4$ supersymmetric Yang-Mills theory as an
effective field theory on parallel D3-branes.

\end{abstract}
\end{titlepage}

\section{Introduction}

The recent developments in non-perturbative string theory have provided
new tools to investigate non-perturbative features of field theory.
Many supersymmetric gauge theories can be studied as effective
field theories on branes.
In this picture, the non-perturbative features of the
supersymmetric gauge theories are tightly related to those of string
theory.
A BPS state of string theory has a correspondence to a BPS state of the
effective field theory on the brane.

The 3+1 dimensional
${\cal N}=4$ $SU(N)$ supersymmetric
Yang-Mills (SYM) theory broken spontaneously to $U(1)^{N-1}$
can be studied as an effective field theory on $N$ parallel
D3-branes \cite{WIT,TSE}.
Since a D3-brane is an invariant object under the $SL(2,Z)$ duality
transformation of IIB string theory, the $SL(2,Z)$ duality
implies the $SL(2,Z)$ duality symmetry \cite{MON}
of the ${\cal N}=4$ SYM theory.
A fundamental string between different  D3-branes corresponds to
a W-boson of the SYM theory, which is a BPS state preserving
$\frac12$ of the supersymmetries on the D3-branes.
By performing the $SL(2,Z)$ duality transformation on the
configurations, a $(p,q)$ string should correspond to a BPS state of
the field theory with electric charge
$p$ and magnetic charge $q$, where $p$ and $q$ are relatively prime
integers.
The state with  $p=0$ and $q=1$ is a monopole, while
states with $p\ne 0$ and $q=1$ are dyons. Their field configurations
are explicitly known in the form of the Prasad-Sommerfield solution
\cite{PS}.
It is a difficult question whether the other dyon states exist as
well in the field theory. In the simplest case of $SU(2)$ with $q=2$,
the existence was shown by the quantization of the collective modes
of the two monopole solutions \cite{SENMON}.

Recently, BPS configurations of IIB strings with 3-string junctions
have been attracting attentions.
It was conjectured originally
in \cite{SHW} that a 3-string junction would exist
under both the conditions that the forces from the
three strings should balance and that the $(p,q)$ charges of the
three strings should conserve at the junction.
The BPS feature of the 3-string junction has been shown recently by
several authors in both the string picture \cite{DAS,SENNET} and the
M-theory picture \cite{KROLEE}.
The relations to the BPS states of field theory have been discussed
in several contexts \cite{ZWI}-\cite{IMA}.
The U-dual of the Hanany-Witten effect \cite{HANWIT} implies the
existence of a certain series of 3-string junctions,
and they were used in the description of the exceptional
Lie groups \cite{ZWI}. On the other hand,
requiring only the above two conditions of 3-string junction,
more BPS string states were observed than those expected in
field theory \cite{FAY}.
The selection rule of proper BPS string states has not been
established yet.

Thus it would be interesting to show explicitly the existence of
the field theory BPS states corresponding to some BPS string states
with 3-string junctions in a simple context.
Since a IIB string with any $(p,q)$ charge can end on a D3-brane,
there would be a BPS state of ${\cal N}=4$ SYM theory corresponding to a
3-string junction connecting three different D3-branes.
These BPS states preserve only $\frac14$ of the D3-brane world volume
supersymmetry, and hence these states are different from the BPS
states mentioned above.
An evidence for the existence of such BPS states in ${\cal N}=4$ SYM
theory was recently obtained in \cite{BER}.
The author argued that such a BPS state has non-parallel electric
and magnetic charges in the world volume theory, and showed that,
under the assumption of BPS saturation,
the mass of such a state agrees with the mass obtained from the
IIB string picture.

In this paper, we find spherically symmetric regular
solutions to the Bogomol'nyi
equations in the cases of non-parallel electric and magnetic charges.
Our general solution corresponds to the three-pronged strings
carrying charges $(a,2)$, $(b,0)$ and $(-a-b,-2)$,
where $a$ and $b$ take arbitrary real values.

\section{Bogomol'nyi equations and the asymptotic behavior}
\label{sec:2}

Since we are interested in the 3-string junction, let us consider the
simplest case of three D3-brane system, and hence
the 3+1 dimensional ${\cal N}=4$ SYM system with gauge group $SU(3)$.
Among the six scalar fields describing the transverse coordinates
of the D3-branes, we keep only the two, $\X$ and $\Y$,
corresponding to the two dimensional plane on which the three-pronged
strings lie.
Therefore, the bosonic part of the hamiltonian of the system reads
\begin{eqnarray}
H=\int d^3x \Half\tr\left\{
\left(\B_i\right)^2 + \left(\E_i\right)^2
+\left(D_i \X\right)^2 +\left(D_i \Y\right)^2
+\left(D_0 \X\right)^2 +\left(D_0 \Y\right)^2
-\com{X}{Y}^2\right\} ,
\label{eq:H}
\end{eqnarray}
where $\B_i=\half\epsilon_{ijk}F_{jk}$ and $\E_i=F_{0i}$ are
the magnetic and electric fields, respectively, and the covariant
derivative is defined by
$D_\mu\X=\p_\mu\X -i\left[\A_\mu,\X\right]$.
We have put the Yang-Mills coupling constant equal to one, and shall
consider the case of vanishing vacuum theta angle.
The BPS saturation condition is derived by introducing an angle
$\theta$ as \cite{FH}
\begin{eqnarray}
&&H=\int d^3x \Half\tr\left\{
\left(\E_i\cos\theta - \B_i\sin\theta - D_i\X\right)^2
+\left(\B_i\cos\theta + \E_i\sin\theta - D_i\Y\right)^2
\right.
\nn\\
&&\qquad\left.
+\left(D_0 \X\right)^2 +\left(D_0 \Y\right)^2 -\com{X}{Y}^2
\right\}
+\left(Q_\X + M_\Y\right)\cos\theta
+\left(Q_\Y - M_\X\right)\sin\theta
\nn\\
&&\phantom{H}
\ge \sqrt{\left(Q_\X + M_\Y\right)^2 + \left(Q_\Y - M_\X\right)^2} ,
\label{eq:BPSbound}
\end{eqnarray}
where $M_{\X,\Y}$ and $Q_{\X,\Y}$ are defined by
\begin{equation}
M_\X= \int\! d^3x \tr\left(\B_i D_i \X\right)
=\!\! \int\limits_{\r\to\infty}\!\! dS_i \tr\left(\B_i \X\right) ,
\quad
Q_\X= \int\! d^3x \tr\left(\E_i D_i \X\right)
=\!\! \int\limits_{\r\to\infty}\!\! dS_i \tr\left(\E_i \X\right) .
\label{eq:M&Q}
\end{equation}
The lower bound in (\ref{eq:BPSbound}) is saturated when the following
conditions hold:
\begin{eqnarray}
&&D_i\X = \E_i\cos\theta - \B_i\sin\theta , \label{eq:DX=B+E}\\
&&D_i\Y = \B_i\cos\theta + \E_i\sin\theta , \label{eq:DY=B+E}\\
&&D_0\X = D_0\Y = 0 , \label{eq:D0X=0}\\
&&\com{X}{Y}=0 , \label{eq:Dflat}
\end{eqnarray}
and the corresponding angle $\theta$ is given by
\begin{equation}
\tan\theta = \left(Q_\Y - M_\X\right)/\left(Q_\X + M_\Y\right) .
\label{eq:theta}
\end{equation}
Besides the four equations (\ref{eq:DX=B+E})--(\ref{eq:Dflat}),
we have to impose the Gauss law,
\begin{equation}
D_i\E_i = 0 , \label{eq:DE=0}
\end{equation}
since we used it in converting the volume integration of $Q_{\X,\Y}$
(\ref{eq:M&Q}) into the surface one.
Note that eq.\ (\ref{eq:theta}) is an automatic consequence of the two
equations (\ref{eq:DX=B+E}) and (\ref{eq:DY=B+E}) and need not be
imposed separately.

\begin{figure}[htdp]
\begin{center}
\begin{minipage}{70mm}
\begin{center}
\leavevmode
\epsfxsize=60mm
\epsfbox{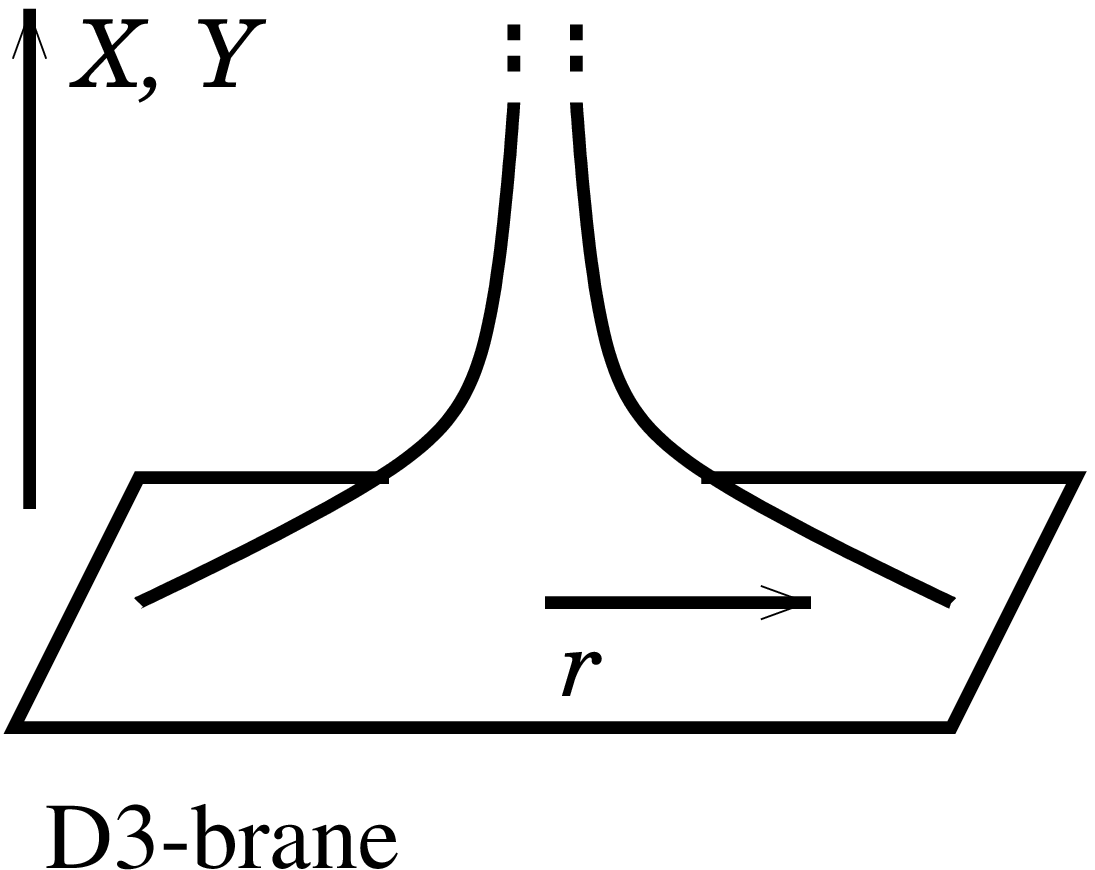}
\caption{``Tube-like'' configuration of D3-brane surface
  representing a string.}
\label{fig:tube}
\end{center}
\end{minipage}
\hspace{5mm}
\begin{minipage}{80mm}
\begin{center}
\leavevmode
\epsfxsize=70mm
\epsfysize=50mm
\epsfbox{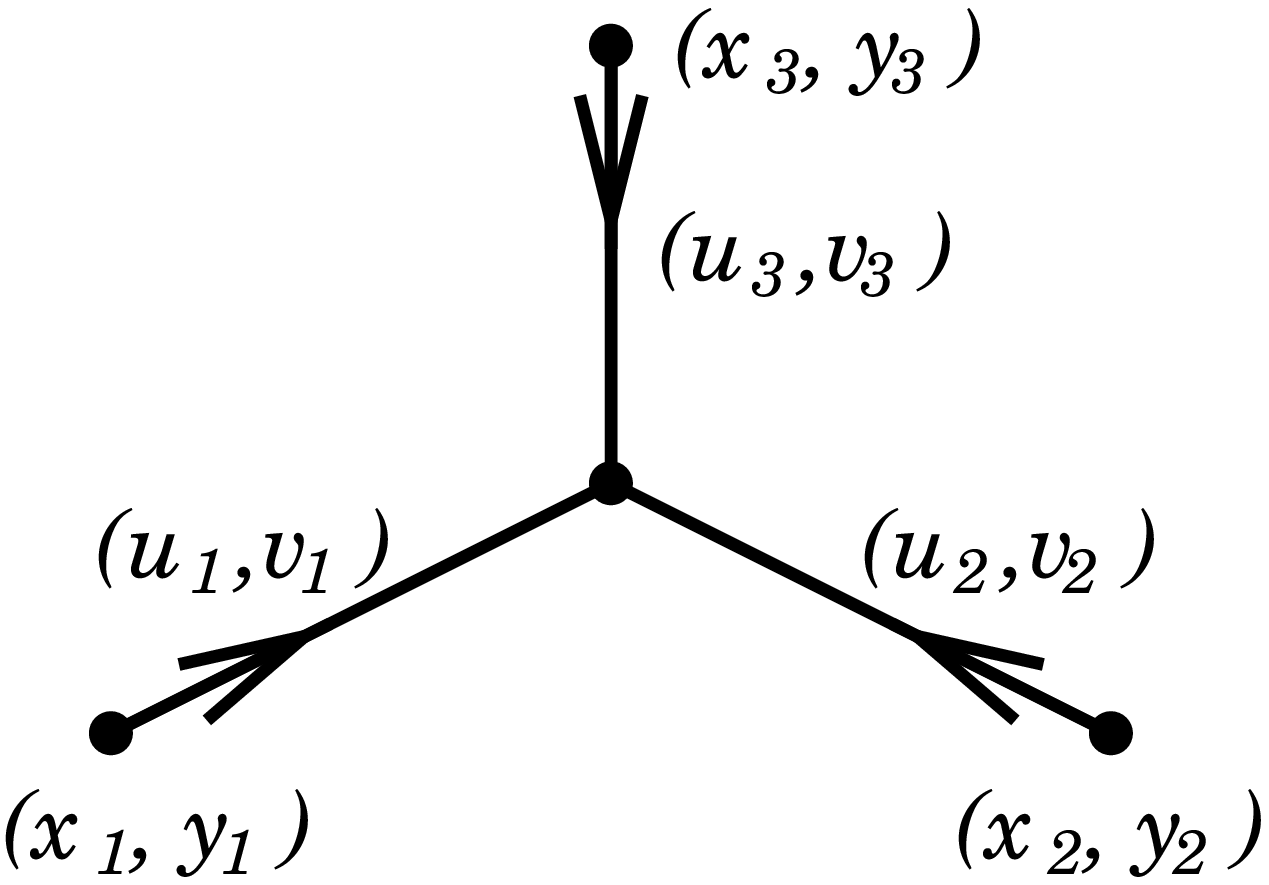}
\caption{3-string junction connecting three D3-branes.}
\label{fig:3jun}
\end{center}
\end{minipage}
\end{center}
\end{figure}

Suppose that we have a static solution
$(\A_\mu(\bmr),\X(\bmr),\Y(\bmr))$
to the equations (\ref{eq:DX=B+E})--(\ref{eq:Dflat}) and
(\ref{eq:DE=0}), and that their asymptotic ($\r\to\infty$) forms are,
after a suitable gauge transformation, given (locally) as follows:
\begin{eqnarray}
&&\X \sim \diag(\x_1,\x_2,\x_3)+\frac{1}{2r}\diag(\u_1,\u_2,\u_3) ,
\label{eq:asX}\\
&&\Y \sim \diag(\y_1,\y_2,\y_3)+\frac{1}{2r}\diag(\v_1,\v_2,\v_3) ,
\label{eq:asY}\\
&&\E_i \sim \frac{\hr_i}{ r^2}\;\Half\diag(\e_1,\e_2,\e_3) ,
\label{eq:asE}\\
&&\B_i \sim \frac{\hr_i}{ r^2}\;\Half\diag(\g_1,\g_2,\g_3) ,
\label{eq:asB}
\end{eqnarray}
with $\hat{\bmr}\equiv \bmr/r$.
This solution represents a configuration of three D3-branes
$a=1,2,3$ at transverse coordinates $(\x_a,\y_a)$ from which a string
with magnetic and electric charges $(\e_a,\g_a)$ are emerging in the
direction $(\u_a,\v_a)$.
This is because the eigenvalues of the scalars $(\X,\Y)$ are
interpreted as the transverse coordinates of the D3-branes and
the ``tube-like'' part of the D3-brane surface (corresponding to small
$\r$) can be regarded as a string \cite{CH}
(see fig.\ \ref{fig:tube}).
Of course, $(\u_a,\v_a)$ are not arbitrary but should be determined in
terms of other parameters of the solution.
In fact, this relation can be obtained by the analysis in the
asymptotic region without concretely solving the differential
equations:
Multiplying eqs.\ (\ref{eq:DX=B+E}) and (\ref{eq:DY=B+E}) with either
$\E_i$ or $\B_i$, we obtain a gauge invariant equation
\begin{equation}
\p_i\tr\left(\E_i\X\right)
=\tr\left(\E_i\right)^2\cos\theta
- \tr\left(\E_i\B_i\right)\sin\theta ,
\label{eq:gaugeinveq}
\end{equation}
and three others. Plugging the asymptotic expressions
(\ref{eq:asX})--(\ref{eq:asB}) into these equations and comparing
the $O\left(1/r^4\right)$ terms,\footnote{
\label{footnote}
Here, we have to assume that the $O\left(1/r^3\right)$ terms are
missing from $\B_i$ and $\E_i$. This assumption will prove valid in
the concrete solutions given later.
}
we get the known relation between the directions of the strings and
their charges \cite{SENNET}:
\begin{equation}
\pmatrix{\u_a\cr \v_a}
= -\pmatrix{\cos\theta & -\sin\theta\cr
             \sin\theta &  \cos\theta}\pmatrix{\e_a\cr \g_a} .
\label{eq:(u,v)}
\end{equation}

The above observations in the asymptotic region lead to the
standard picture of the 3-string junction
(see fig.\ \ref{fig:3jun}):
We can show that the three straight lines starting at $(\x_a,\y_a)$
in the direction $(\u_a,\v_a)$ meet at a common point.
Hence, these straight lines can be regarded as IIB strings
forming a junction.
In fact, the force balance relation,
$\sum_a T_a(\u_a,\v_a)(\u_a^2+\v_a^2)^{-1/2}=0$, holds by taking
$T_a=\left(\e_a^2+\g_a^2\right)^{1/2}$ as the string tension.
Furthermore, the sum of the masses of the three strings,
$\sum_i T_i\ell_i$, with $\ell_a$ being the length of the $a$-th
straight line, coincides with the BPS bound (\ref{eq:BPSbound}) of the
SYM hamiltonian \cite{BER}.
However, we shall see that the configurations of the 3-string
junctions obtained from the eigenvalues of the scalars $(\X,\Y)$
of the classical solutions of SYM constructed
below have more complicated structures than the above string picture.

\section{Solutions}
\label{sec:3}

Since the angle $\theta$ can be absorbed by the rotation in the
$(\X,\Y)$ plane, we restrict ourselves to the case with $\theta=0$.
Then, equations to be solved are
\begin{eqnarray}
&&D_i \X = \E_i , \label{eq:DX=E}\\
&&D_i \Y = \B_i , \label{eq:DY=B}
\end{eqnarray}
as well as (\ref{eq:D0X=0}), (\ref{eq:Dflat}) and (\ref{eq:DE=0}).
Our strategy for the construction of the solutions is as follows.
First, we prepare a (monopole) solution
$\left(\A_i(\bmr),\Y(\bmr)\right)$ to eq.\ (\ref{eq:DY=B}).
Then, eq.\ (\ref{eq:DX=E}) is automatically satisfied by putting
$A_0(\bmr)= -\X(\bmr)$, while eq.\ (\ref{eq:D0X=0}) holds due to
eq.\ (\ref{eq:Dflat}) and the time-independence of our solution.
Therefore, we have only to solve eq.\ (\ref{eq:DE=0}), i.e.,
\begin{equation}
D_iD_i \X=0 ,
\label{eq:DDY=0}
\end{equation}
under the D-flatness condition (\ref{eq:Dflat}).

The monopole solutions to eq.\ (\ref{eq:DY=B}) for a gauge group
$SU(N)$ were discussed by many people. Among them, we adopt the
solutions given by refs.\ \cite{BW,WB} constructed on the
basis of the general formalism of spherically symmetric solutions of
ref.\ \cite{WG}.
Let us first recapitulate the elements of ref.\ \cite{WB}
necessary for our construction.

Our solutions $\left(A_i, \X, \Y\right)$ are assumed to be spherically
symmetric with respect to $SO(3)$ generator $\bm{J}$, i.e., they
satisfy
\begin{eqnarray}
\com{J_i}{A_j}=i\epsilon_{ijk}A_k ,
\qquad
\com{J_i}{\X}=\com{J_i}{\Y}=0 ,
\label{eq:sphsymm}
\end{eqnarray}
and the present $\bm{J}$ is given as
\begin{equation}
\bm{J}=\bm{L} + \bm{T} ,
\label{eq:J=L+T}
\end{equation}
where $\bm{L}= -i\bmr\times\bm{\nabla}$ is the generator of the space
rotation and $\bm{T}$ is the maximal $SU(2)$ embedding in $SU(3)$ with
$T_3 =\diag\left(1,0,-1\right)$.
The monopole solution of \cite{WG} takes the following form for the
vector potential:
\begin{equation}
\bm{A}(\bmr)=\left(\bm{M}(r,\hbmr) - \bm{T}\right)\times\hbmr/r ,
\label{eq:A=}
\end{equation}
where the Lie algebra valued function $M_i$ should satisfy
the spherical symmetry condition,
\begin{equation}
\com{J_i}{M_j}=i\epsilon_{ijk}M_k .
\label{eq:[J,M]}
\end{equation}
Various formulas are derived by using the expression
$\bm{\nabla}=\hbmr\pdrv{}{r}-(i/r)\hbmr\times\bm{L}$ for the space
derivative as well as the spherical symmetry properties,
eqs.\ (\ref{eq:sphsymm}) and (\ref{eq:[J,M]}).
We list the three necessary for our purpose:
\begin{eqnarray}
&&\bm{D}\Y=\hbmr \Y' + \frac{i}{r}\,\hbmr\times\com{\bm{M}}{\Y} ,
\label{eq:DX2}\\
&&B_i = -\frac{i}{r^2}\hr_i\hr_j\left(
\Half\epsilon_{jk\ell}\com{M_k}{M_\ell} - iT_j\right)
- \frac{1}{r}\left(\hbmr\times\left(
\hbmr\times\bm{M}'\right)\right)_i ,
\label{eq:B}\\
&&D_iD_i\X = \X'' + \frac{2}{r}\X'
- \frac{1}{r^2}\left(\delta_{ij}-\hr_i\hr_j\right)
\com{M_i}{\com{M_j}{\X}} ,
\label{eq:DDY}
\end{eqnarray}
where the prime denotes the differentiation $\pdrv{}{r}$.

Due to spherical symmetry we have only to construct solutions on the
positive $z$-axis. The monopole equation (\ref{eq:DY=B}) on the
$z$-axis is reduced to
\begin{eqnarray}
&&r^2 \Y'= \Half\com{M_+}{M_-} - T_3 ,
\label{eq:D3X=B3}\\
&&M'_{\pm} = \mp\com{M_\pm}{\Y} ,
\label{eq:DpmX=Bpm}
\end{eqnarray}
with $M_\pm\equiv M_1\pm i M_2$. The solution to eqs.\ (\ref{eq:D3X=B3})
and (\ref{eq:DpmX=Bpm}) given in ref.\ \cite{WB} for the $SU(3)$ case
is as follows. The matrices $X$ and $M_\pm$ are expressed as
\begin{eqnarray}
&&\Y=\diag\left(\Y_a\right)
=\Half\diag\left(\phi_1,\phi_2-\phi_1,-\phi_2\right) ,
\label{eq:X}\\
&&M_+=\pmatrix{0 & a_1 & 0   \cr
             0 & 0   & a_2 \cr
             0 & 0   & 0} ,
\quad M_-=\left(M_+\right)^T ,
\label{eq:M+}
\end{eqnarray}
in terms of $a_m$ and $\phi_m$, which are further given in terms of
two functions $(Q_1,Q_2)$ as
\begin{equation}
a_m = \frac{r}{Q_m}\left(2\, Q_{3-m}\right)^{1/2} ,
\qquad
\phi_m = -\Drv{\ln Q_m}{r} + \frac{2}{r} .
\label{eq:phi_m}
\end{equation}
Eq.\ (\ref{eq:DpmX=Bpm}) is an automatic consequence of the
expressions (\ref{eq:phi_m}), while eq.\
(\ref{eq:D3X=B3}) now becomes
\begin{equation}
\left(Q'_m\right)^2 - Q_m Q''_m = Q_{3-m} .
\label{eq:eqQ}
\end{equation}
In ref.\ \cite{WB} they found the following solution to
(\ref{eq:eqQ}):
\begin{equation}
Q_1(\r)=\Half\sum_{a=1,2,3}
\exp\left(-2\y_a r\right)
\prod_{b\,(\ne a)}\left(\y_a - \y_b\right)^{-1} ,
\quad
Q_2(\r)=Q_1(-\r) .
\label{eq:solQ}
\end{equation}
Here, $\y_a$ ($\sum_a\y_a=0$) are the constants characterizing the
solution, and they are nothing but $\y_a$ in the asymptotic expression
(\ref{eq:asY}) of $\Y$ if the condition $y_1<y_2<y_3$ is satisfied.
The magnetic charges $\g_a$ of the present solution are
\begin{equation}
(\g_1,\g_2,\g_3)=(-2, 0, 2) .
\label{eq:g}
\end{equation}

Having prepared the monopole solution $(\A_i,\Y)$, our task is to
solve eq.\ (\ref{eq:DDY=0}) for the scalar $\X$, which,
using eq.\ (\ref{eq:DDY}), becomes on the $z$-axis
\begin{eqnarray}
\X'' + \frac{2}{r}\X' - \frac{1}{2r^2}\Bigl(
\com{M_+}{\com{M_-}{\X}} + \com{M_-}{\com{M_+}{\X}}\Bigr)=0.
\label{eq:Y''}
\end{eqnarray}
Since we have adopted a diagonal form (\ref{eq:X}) for $\Y$,
eq.\ (\ref{eq:Dflat}) implies that $\X$ is also diagonal:
\begin{eqnarray}
\label{eq:diagY}
\X = \diag \left(\X_a\right)
   = \frac{1}{4r}\diag\left(\nr +\pr, -2\pr,-\nr+\pr\right) .
\end{eqnarray}
Then, eq.\ (\ref{eq:Y''}) for $\varphi_\pm$ reads
\begin{eqnarray}
  \label{eq:diff}
\varphi''_\pm
-\left(\frac{Q_2}{Q_1^2}\pm\frac{Q_1}{Q_2^2}\right)\varphi_+
-3\left(\frac{Q_2}{Q_1^2}\mp\frac{Q_1}{Q_2^2}\right)\varphi_-
=0
\end{eqnarray}
The differential equation (\ref{eq:diff}) can be solved both
numerically and analytically (we shall present exact solutions in
Sec.\ \ref{sec:3.2}).
However, let us first carry out the analysis of the solutions near the
origin and infinity. This is useful for the understanding of the
solutions given later.

\subsection{Behavior of the solution near $\r=0$ and $\infty$}
\label{sec:3.1}

First, we shall consider (\ref{eq:diff}) near the origin $\r=0$.
Since $Q_m(r)$ is Taylor-expanded around $\r=0$ as
\begin{eqnarray}
\label{eq:Q12r0}
Q_m = \r^2 -\frac{1}{3}\sum_{a>b}\y_a\y_b\,\r^4
+(-)^m \frac{2}{15}\y_1\y_2\y_3\,\r^5 + O(\r^6) ,
\end{eqnarray}
eq.\ (\ref{eq:diff}) is approximated near the origin as
\begin{eqnarray}
\label{eq:diffr0}
\nr ''(\r) - \left( \frac{2}{\r^2} + O(1)\right)\nr(\r) = 0 ,
\qquad
\pr ''(\r) - \left( \frac{6}{\r^2} + O(1)\right)\pr(\r) = 0 .
\end{eqnarray}
Each of the differential equations (\ref{eq:diffr0}) has two
independent solutions; one is regular and the other is
singular at $\r=0$. From the physical requirement, we of course have to
choose regular ones which behave as
\begin{eqnarray}
\label{eq:solnpr0}
\nr = \wt{\alpha}\;\r^2 + O(\r^4),\qquad
\pr = \wt{\beta}\;\r^3 +O(\r^5),
\end{eqnarray}
where $(\wt{\alpha},\wt{\beta})$ are arbitrary real constants.
Once the parameter $(\wt{\alpha},\wt{\beta})$ is
fixed, we can successively determine the coefficients of any higher
powers of $\r$ by iterative use of eq.\ (\ref{eq:diff}).
Therefore, taking also into account the parameter $\y_a$ of the
monopole solution, we see that the number of the freedom of our
solutions $(\A_\mu,X,Y)$ is four;
$(\y_1,\y_2,\wt{\alpha},\wt{\beta})$.

{}From eqs.\ (\ref{eq:solnpr0}), (\ref{eq:phi_m}) and (\ref{eq:Q12r0}),
we obtain the following behavior of the eigenvalues
$\left(\X_a(\r),\Y_a(\r)\right)$ near $\r=0$:
\begin{eqnarray}
\label{eq:r0}
(\X_1,\Y_1)\sim -(\X_3,\Y_3)
\sim \left(
\frac{\wt{\alpha}}{4}\, ,\,\frac{1}{3}\sum_{a>b}\y_a\y_b\right)\r ,
\quad
(\X_2,\Y_2)\sim -\left(
\frac{\wt{\beta}}{2}\, ,\,\frac{2}{5}\y_1\y_2\y_3\right)\r^2 .
\end{eqnarray}
Eq.\ (\ref{eq:r0}) implies the followings:
First, since both the scalars $\X$ and $\Y$ representing the D3-brane
transverse coordinates vanish at $\r=0$,
the three D3-brane surface meet at the origin $(\X,\Y)=(0,0)$
of the transverse plane.
Second, among the three D3-branes, the branes 1 and 3 are ``smoothly''
connected at the origin (namely, they have a common tangent at the
junction), while the brane 2 meets with the other two
with an angle.

Next let us consider eq.\ (\ref{eq:diff}) near $\r=\infty$.
Since both the quantities $Q_2/Q_1^2$ and $Q_1/Q_2^2$ which appear in
eq.\ (\ref{eq:diff}) decay exponentially as $\r\to\infty$, the leading
behavior of $\X$ is in fact given by (\ref{eq:asX}), and the next
order terms are exponentially dumping ones.
The other scalar $\Y$ behaves in a similar manner at infinity.
This fact, together with the formulas $\E_z= \X'$ and $\B_z = \Y'$
on the $z$-axis, which are consequence of eqs.\ (\ref{eq:DX=E}),
(\ref{eq:DY=B}) and (\ref{eq:DX2}), implies the validity of the
assumption we made in deriving eq.\ (\ref{eq:(u,v)}):
the $O(1/\r^3)$ terms are missing from $\E$ and $\B$.

\subsection{Exact solutions for $\X$}
\label{sec:3.2}

Though it seems difficult to solve analytically the differential
equation (\ref{eq:diff}) for arbitrary $(\y_a)$,
exact solutions are obtained for the following special values of
$(\y_a)$:
\begin{eqnarray}
(\y_1,\y_2,\y_3)=(-C, 0, C),
\end{eqnarray}
with $C$ being a real and positive constant.
In fact, for this $(\y_a)$ we have
\begin{eqnarray}
\label{eq:Q1Q2}
  Q_1 = Q_2 = \left(\frac{\sinh C\r}{C}\right)^2 ,
\end{eqnarray}
and hence the differential equations (\ref{eq:diff}) become
separated ones:
\begin{eqnarray}
\nr ''(\r) - \frac{2C^2}{\sinh^2 C\r}\nr(\r) = 0,
\qquad
\pr ''(\r) - \frac{6C^2}{\sinh^2 C\r}\pr(\r) = 0.
\label{eq:diff-sep}
\end{eqnarray}
Eqs.\ (\ref{eq:diff-sep}) have the following solutions regular at
$\r=0$:
\begin{eqnarray}
\nr(\r) = \alpha(C\r\coth C\r -1 ),
\qquad
\pr(\r) = \beta\left(
3\coth C\r - C\r \frac{2\cosh^2C\r + 1}{\sinh^2C\r} \right).
\label{eq:solution}
\end{eqnarray}
where $\alpha$ and $\beta$ are arbitrary constants.
These solutions are consistent with (\ref{eq:solnpr0}) by the
identifications
$\wt{\alpha}=\alpha C^2/3$ and $\wt{\beta}=-4\beta C^3/15$.

{}From eq.\ (\ref{eq:solution}), we can read off the following
values for the locations of the D3-branes and the electric and
magnetic charges of the strings:
\begin{eqnarray}
&&(\x_a,\y_a)=\left\{C\left(\frac{\alpha -2\beta}{4},\,-1\right),
\quad
C\Bigl(\beta,\,0\Bigr),
\quad
C\left(-\frac{\alpha +2\beta}{4},\,1\right)
\right\} ,
\label{eq:D3}\\
&&(e_a,g_a) = \left\{
\left(\frac{\alpha -3\beta}{2},\,-2\right),\quad
\Bigl(3\beta,\,0\Bigr),\quad
\left(-\frac{\alpha +3\beta }{2},\,2\right)
\right\} .
\label{eq:charge}
\end{eqnarray}
\begin{figure}
\begin{center}
\leavevmode
\epsfxsize=80mm
\epsfbox{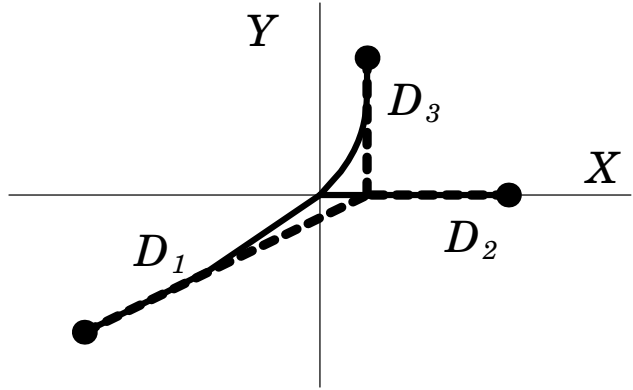}
\caption{The D3-brane configurations (solid lines) and
the three-pronged strings from the IIB picture (dashed lines).}
\label{fig:3}
\end{center}
\end{figure}
In particular, for $(\alpha,\beta)=(-1,1/3)$, the three charges are
$(-1,-2)$, $(1,0)$ and $(0,2)$.
In fig. \ref{fig:3}, we plot
the trajectories of the D3-brane coordinates in course of changing
$r$ by the solid lines.
In course of decreasing $r$, the D3-branes
approach the origin of $(X,Y)$ and meet there at $r=0$, where the gauge
symmetry is restored.
The trajectory of the brane $D_2$ is just a straight line.
This is a general feature of our exact solutions and comes from the boundary
condition $\y_2=0$.
We obtain a bending trajectory of $D_2$ for a general case
of $\y_2\ne 0$, in which we solved the equations only numerically.

The branes $D_1$ and $D_3$ connect smoothly to each other at the
origin, as discussed in Sec.\ \ref{sec:3.1}.
Noticing the fact that the brane $D_2$ has no magnetic
charge\footnote{We mean the charge $(e_a,g_a)$ by the electric and
magnetic charges of the brane $D_a$.}
while the branes $D_1$ and $D_3$ have non-zero magnetic charges,
this might come from our technical preference that we describe the BPS
states by the classical treatment of the SYM theory,
i.e.\ electrically.
Then the interpretation of the trajectories might be that
the branes $D_1$ and $D_3$ are the two parts of one very heavy smooth
magnetic object pulled and bent by the light electric brane $D_2$.

In fig.\ \ref{fig:3}, we also draw three dashed straight lines
tangent to the D3-brane trajectories at $r= \infty$.
They meet at one point. As discussed in Sec.\ \ref{sec:2},
this configuration agrees
with the three-pronged strings in the IIB string picture.

\section{Summary and Discussions}

We have obtained BPS saturated
spherically symmetric regular configurations in
${\cal N}=4$ $SU(3)$ SYM theory carrying non-parallel electric
and magnetic charges $(a,2)$, $(b,0)$ and $(-a-b,-2)$, where $a$ and
$b$ take arbitrary real values.
Regarding the ${\cal N}=4$ SYM theory as an effective field theory on
parallel D3-branes, the solutions correspond to 3-string
junctions connecting three different D3-branes.
Assuming the quantization of the electric
charges $a$ and $b$ and the $SL(2,Z)$ duality symmetry, our solutions
imply, in general,  the existence of the junctions of the three IIB strings
carrying the two-form charges
$(p,q)$, $(lr,ls)$ and $(-p-lr,-q-ls)$, respectively,
where $l,p,q,r,s$ are integers satisfying $ps-qr=2$.

As discussed in Sec.\ \ref{sec:2},
the asymptotic behavior at $r\sim\infty$ of our solutions leads to
the configuration of three-pronged strings in the IIB string
picture, i.e. three straight strings meet at one point and
their forces balance.
This would be a non-trivial consistency check of our approach.
On the other hand, the behavior of our solution at finite $r$ is
quite different from the above IIB string picture.
The trajectories of the D3-branes bend non-trivially, and
the two magnetically
charged\footnote{See the previous footnote.}
D3-branes connect smoothly to each other.
As discussed in Sec.\ \ref{sec:3.2},
this might be due to the fact that we did not
treat the problem in an $SL(2,Z)$ invariant way, but obtained
the BPS configurations by the classical treatment of the SYM theory,
i.e.\ electrically.

Another difference between our solutions and the IIB string picture
is the number of degrees of freedom.
Now let us count the number of degrees of freedom of a 3-string
junction in the string picture.
As for the charges, the three magnetic charges are fixed to
(\ref{eq:g}) from the beginning, but
we have the freedom of two electric charges
(the other electric charge is determined by the charge conservation).
Then, since the string tensions are determined by the charges,
the relative directions of the
three strings are determined by the force balance condition.
We have the freedom to choose the lengths of each string.
We should not count the freedom of rotating or
shifting parallelly the 3-string junction in the two-dimensional
plane, because we fixed these degrees of freedom in solving the
equations. We absorbed the angle $\theta$ in Sec.\ \ref{sec:3}, and the
center of the three D3-branes are fixed by the tracelessness of the
adjoint scalar fields.
Thus we have in total five degrees of freedom in the string picture.
This does not agree with four, the number of degrees of freedom of our
solutions obtained in Sec.\ \ref{sec:3.1}.

This difference might again come from the limitation of the electric
description of the 3-string junction. Since electrically charged
objects are the fundamental
degrees of freedom themselves in the electric
description, the introduction of a bare electric
charge would necessarily cause the problem of singularities in the
solutions. Nevertheless, in our solutions, we have one D3-brane
($D_2$ in fig.\ \ref{fig:3}) which is charged only electrically.
However, we may have the possibility
that, if we had the freedom to introduce a bare electric charge, the
number of the degrees of freedom would increase from four to five.

In fact, the difficulty of a bare electric charge appears in a
different way in our construction of the solutions.
We took the maximal embedding of $SU(2)$ to $SU(3)$ in this paper.
Because of this, the magnetic charge is 2 and not 1.
Taking the minimal embedding of $SU(2)$ does not work. It turns out
that, to obtain non-parallel electric and magnetic charges, we have
a singularity at the origin, which is the bare source of the electric
field.

We tried another way to introduce a unit magnetic charge. In the
degenerate case of $y_1=y_2$, the magnetic charges of the D3-branes
become $(1,1,-2)$ \cite{BW,WB}.
But in this case the asymptotic behavior at $r\sim\infty$
of eq.\ (\ref{eq:diff}) changes from the non-degenerate cases,
and the solutions regular at
$r=0$ diverge at $r=\infty$
($\lim_{r\rightarrow\infty} \phi_1(r)=\infty$)
except the case of parallel electric and magnetic charges.
Thus we could not introduce one unit of magnetic charge in our
solutions.

Since the monopole solutions for the general $SU(N)$ case are also
known, it would be an interesting
extension to find BPS saturated
solutions carrying non-parallel electric and magnetic charges  for
general $SU(N)$ and compare them with the string picture.
This work is now in progress.

\vspace{.5cm}
\noindent
{\large\bf Acknowledgments}\\[.2cm]
We would like to thank Y.\ Imamura, I.\ Kishimoto, H.\ Kunitomo and
S.\ Sugimoto for valuable discussions.

\newcommand{\J}[4]{{\sl #1} {\bf #2} (#3) #4}
\newcommand{\andJ}[3]{{\bf #1} (#2) #3}
\newcommand{\AP}{Ann.\ Phys.\ (N.Y.)}
\newcommand{\MPL}{Mod.\ Phys.\ Lett.}
\newcommand{\NP}{Nucl.\ Phys.}
\newcommand{\PL}{Phys.\ Lett.}
\newcommand{\PR}{Phys.\ Rev.}
\newcommand{\PRL}{Phys.\ Rev.\ Lett.}
\newcommand{\PTP}{Prog.\ Theor.\ Phys.}

\end{document}